\documentclass[lettersize,journal]{IEEEtran}
\usepackage{amsmath,amsfonts}
\usepackage{algorithmic}
\usepackage{algorithm}
\usepackage{array}
\usepackage{textcomp}
\usepackage{stfloats}
\usepackage{url}
\usepackage{verbatim}
\usepackage{graphicx}

\usepackage{siunitx}
\usepackage{subcaption}
\usepackage{enumerate}

\newcommand{\eref}[1]{Eq.\,(\ref{#1})}

\newcommand{\figref}[1]{Fig.\,\ref{#1}}
\newcommand{\Figref}[1]{Figure\,\ref{#1}}
\newcommand{\tabref}[1]{Table\,\ref{#1}}
\newcommand{\secref}[1]{Section\,\ref{#1}}

\begin{document}

\title{Screening Curve Method for Economic Analysis \\of {\color{red}Household Solar Energy Self-Consumption}}

\author{Hikaru Hoshino,~\IEEEmembership{Member},~
        Yosuke Irie,~
        Eiko Furutani,~\IEEEmembership{Member}
\thanks{This work was supported in part by JST, ACT-X Grant Number JPMJAX210M, Japan.}
\thanks{The authors are with the Department of Electrical Materials and Engineering, University of Hyogo, 2167 Shosha, Himeji, Hyogo, 671-2280, Japan (email: hoshino@eng.u-hyogo.ac.jp). }%
}

\markboth{}%
{Shell \MakeLowercase{\textit{et al.}}: A Sample Article Using IEEEtran.cls for IEEE Journals}

\IEEEpubid{}

\maketitle

\begin{abstract}
The profitability of {\color{red}solar energy self-consumption in households, the so-called photovoltaic (PV) self-consumption,} is expected to boost the deployment of PV and battery storage systems. This paper develops a novel method for economic analysis of PV self-consumption using battery storage based on an extension of the Screening Curve Method (SCM). The SCM enables quick and intuitive estimation of the least-cost generation mix for a target load curve and has been used for generation planning for bulk power systems. In this paper, we generalize the framework of existing SCM to take into account the intermittent nature of renewable energy sources and apply it to the problem of optimal sizing of PV and battery storage systems for a household. Numerical studies are provided to verify the estimation accuracy of the proposed SCM and to illustrate its effectiveness in a sensitivity analysis, owing to its ability to show intuitive plots of cost curves for researchers or policy-makers to understand the reasons behind the optimization results. 
\end{abstract}


\section{Introduction}

%
\IEEEPARstart{T}{he} installation capacity of solar photovoltaics (PV) in the world has increased over the past decades owing to the use of support policies such as  feed-in tariffs \cite{timilsina12,poruschi18}.
Recently, the price of PV has declined dramatically, and the grid parity has been reached in several countries \cite{iea-pvps22}, which means that PV can produce electricity at a price below the price of electricity purchased from the grid. 
{\color{red}In these regions or countries, not only selling electricity to the grid but also self-consumption of PV-produced electricity can bring economic benefit to the consumer. 
Because of this, 
the deployment of PV based on 
household solar energy self-consumption,
which is called as PV self-consumption \cite{luthander15}, 
has attracted great interest.} 
As one of the policies based on the concept of PV self-consumption,
many countries introduced net-metering or net-billing mechanisms \cite{ordonez22}, where the surplus electricity fed into the grid is rewarded with credits that can be applied to offset consumption.
However, there have been intense debates on net‑metering policies regarding  recovery of network costs and equity of cost allocation \cite{felder14,schittekatte18}, and several setbacks and cancellations in the application of these policies have occurred \cite{iea-pvps22}. 
Furthermore, various new self-consumption schemes, such as collective or distributed self‑consumption \cite{ines20,dadamo22}, are proposed or implemented 
in several countries,  
and thus appropriate regulatory frameworks based on the concept of self-consumption remain to be discussed.  

%
To contribute to such discussions, economic analysis for evaluating the profitability of PV self-consumption has been extensively studied. 
Existing methods can be categorized into optimization and simulation
methods, depending on whether 
the capacities of PV and battery storage
are optimization variables or simulated as exogenous parameters \cite{han22}.
Since simulation-based methods might underestimate the economic
value of the investment \cite{zhang17:pv}, it is important
to correctly determine the sizes of the PV and battery storage according to the customers’ load
profiles, and thus optimization methods are often used  \cite{nyholm16,schopfer18,cervantes18}. 
While these methods can incorporate constraints to describe various situations, it takes hours or days to obtain optimization results.  
Therefore, only a limited number of scenarios (input data and parameters) are likely to be analyzed 
to provide inadequate information to infer a trend for inputs.
Besides, these methods gives the optimization results like a ``black box", and limited intuition can be gained from the obtained results. 
{\color{red}
Since there is a high uncertainty in the estimates of parameters such as prices and costs,
the above two issues of computational time and the interpretability of the results make it difficult for policy-makers and regulators to learn from
these optimization studies.}

%
The Screening Curve Method (SCM) is an intuitive and quick method 
for estimating 
the least-cost generation mix, 
and it is a good tool to provide answers to the above issues.
SCM was first proposed in the 1960s \cite{phillips69,lee78}.  
It compares the summations of annualized upfront investment costs (fixed costs) and production costs (variable costs) of multiple technologies to determine the optimal mix when no technical constraints are considered. 
In recent years, much effort has been made to develop enhanced versions of SCM to estimate the optimal mix considering technical details. 
The start-up cost, thermal cycling, and other miscellaneous short-term operations of thermal generation are modeled in \cite{batlle13,zhang15}. 
Furthermore, ancillary services are considered in \cite{zhang15:ancillary}, hydro scheduling is modeled in \cite{staffell16}, existing capacities are considered in \cite{zhang17,guner18}, and planned outages are considered in \cite{zhang19}. 
All of these SCMs have been developed for estimating the optimal capacity mix in a bulk power system. 

\IEEEpubidadjcol

%
The purpose of this paper is to develop an SCM for economic analysis of photovoltaic self-consumption to estimate optimal sizes of PV and battery storage systems at the demand side. 
It has been known that the framework of SCM can be applied to the optimal sizing of various sources to satisfy load demands \cite{stoft02,priya14}. 
However, there has been no existing SCM that can take into account the intermittent nature of renewable energy sources. 
In this paper, to consider the fact that the amount of electricity generated by PV depends on the solar irradiation at each time, we newly adopt a time-varying scheme to decompose the load curve, whereas time-invariant load slices have been used in existing SCMs to decompose the original optimization problem into sub-problems. 
It is shown that the proposed SCM can be used to estimate the optimal solution of a Linear Programming (LP) problem. 
While an estimation algorithm has been proposed in \cite{ru14} for the upper bound of the battery size, estimation of the optimal value of the battery size as well as the PV size has not been achieved. 

%
A preliminary version of the proposed SCM was presented in \cite{hoshino21}, but periodicity in the input data was assumed on a time scale of 24 hours. 
The proposed SCM that can handle arbitrary shapes and lengths of data has been partly presented in the conference publication \cite{hoshino22:nolta}. 
This paper adds the following important contributions:
the proposed algorithm is refined, and an analytical expression for estimating the economic benefit of battery is newly introduced; 
a systematic verification study is presented by comparing the results with LP optimal solutions under various settings of parameters; 
and the effectiveness of the proposed SCM is examined in terms of computational time.  

%
This paper is structured as follows. 
\secref{sec:lp} introduces the optimal sizing problem for PV and battery storage systems studied in this paper. 
The proposed SCM is presented in \secref{sec:scm} after a brief review of the existing SCMs. 
\secref{sec:verification} provides a verification study for the estimation accuracy of the proposed SCM under various settings of parameters. 
\secref{sec:discussion} discusses the advantages and disadvantages of the proposed SCM, and \secref{sec:conclusion} concludes this paper.

\section{Optimal Sizing Problem} \label{sec:lp}

The proposed SCM is intended to estimate the optimal solution of the LP problem introduced in this section. 
The problem formulation follows from \cite{schittekatte18} and is based on the following assumptions: 
\begin{itemize}
    \item The sizes of PV and battery storage systems are continuous decision variables, while several discrete sizes would be the options in practice. \item The electricity generated by a unit capacity of PV is proportional to the solar irradiation, and the effect of inverter clipping due to oversizing of a solar array is neglected. 
    \item The profiles of load demand and solar irradiation are completely known a priori to optimize the operation of the battery storage system. 
    \item The electricity retail price per unit of energy consumed is fixed, and other practical tariff structures such as three-part tariff are not considered. 
\end{itemize}

\begin{figure}[!t]
 \centering
 \vspace{2mm}
 \includegraphics[width=\linewidth]{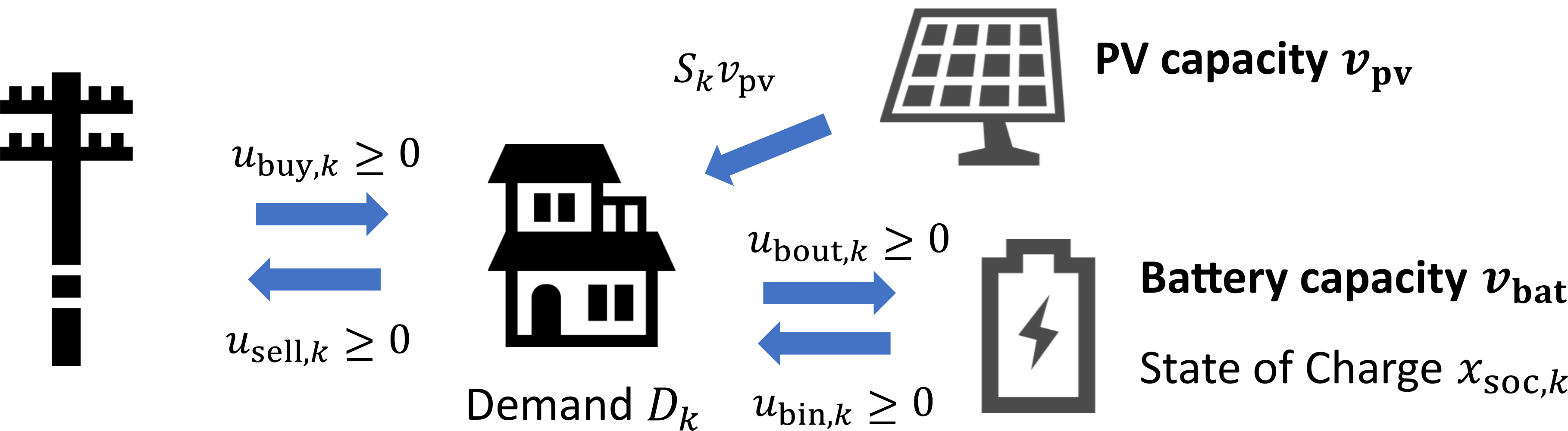} 
 \caption{Optimal PV and battery sizing problem} \label{fig:optimal_sizing}
\end{figure}

The optimal sizes of PV and battery are determined with the aim of minimizing the annualized electricity-related costs of a household while satisfying the load demand at each time. 
This problem can be viewed as a multi-scale decision-making problem, where the decision variables include both the long-term decisions on the installation capacities $v_\mathrm{pv}$ of PV {\color{red}(in $\si{kW}$)} and $v_\mathrm{bat}$ of battery {\color{red}(in $\si{kWh}$)} and the short-term operational decisions on the amounts of electricity purchased from and sold to the grid at each time step $k$, denoted by $u_{\mathrm{buy},k}$ and $u_{\mathrm{sell},k}$ (for $k = 1, \, 2, \,\dots , \, N_\mathrm{t}$), and the amounts of charging and discharging of battery, denoted by $u_{\mathrm{bin},k}$ and $u_{\mathrm{bout},k}$, respectively, where $N_\mathrm{t}$ stands for the number of time steps. 
\Figref{fig:optimal_sizing} shows the overview of the PV and battery sizing problem considered in this paper. 
The meaning of parameters and these typical values are listed in \tabref{tab:parameters}.

\begin{table}[t]%
\footnotesize
\caption{Parameters of the optimization problem}
\label{tab:parameters}
\centering
\begin{minipage}{\linewidth}
\begin{tabular}{c|l|c}
\hline
Symbol & Meaning &  Value \\
\hline
 $C_\mathrm{pv} $ & Annual fixed cost of PV  & $\SI{12000}{yen/(kW \cdot yr)}$ \footnote{It represents annualized cost and corresponds to, e.g., the case where the PV unit price of \SI{200000}{yen/kW} with the lifetime of 25 years, the discount rate of 0\%, and the maintenance and operation cost of \SI{4000}{yen/(kW\cdot yr)}.} \\
 $C_\mathrm{bat} $ &  Annual fixed cost of battery &  $\SI{4400}{yen/(kWh \cdot yr)}$ \footnote{It corresponds to, e.g., the case where the unit price of the battery is \SI{66000}{yen/kWh}, the lifetime of 15 years, and the discount rate of 0\%.} \\
 $P_\mathrm{buy} $ & Price for buying from the grid &  $\SI{26.0}{yen/kWh}$ \\
 $P_\mathrm{sell}$ & Price for selling to the grid &  $\SI{6.0}{yen/kWh}$ \\
 $E_\mathrm{chg} $ & Efficiency of charging  & $0.9$ \\
 $E_\mathrm{dis} $ & Efficiency of discharging & $0.9$ \\
 $E_\mathrm{pv}$   & {\color{red}Performance ratio of PV system} & {\color{red}$ 0.78$} \\
 {\color{red} $G_\mathrm{stc}$} & {\color{red} STC reference irradiance intensity} & {\color{red}$\SI{1000}{W/m^2}$} \\ 
 $S_k$ & Solar irradiation at time $k$ & Given as input data \\
 $D_k$ & Electricity consumption at time $k$ & Given as input data \\ 
 $M_\mathrm{pv}$   & Maximum capacity of PV  & $\SI{10}{kW}$ \\
 $F_\mathrm{anu}$  & Scaling factor for annualization & $365/\text{\# of days analyzed}$ \\
\hline
\end{tabular}
\end{minipage}
\end{table}

The objective function $L$  to be minimized is given as follows:
\begin{align} \label{eq:cost_function}
 L = 
   & F_\mathrm{anu}  \sum_{k=1}^{N_\mathrm{t}}  (P_\mathrm{buy} u_{\mathrm{buy},k} - P_\mathrm{sell} u_{\mathrm{sell},k} )   \notag \\  
   &+ C_\mathrm{pv} v_\mathrm{pv} + C_\mathrm{bat} v_\mathrm{bat} 
\end{align}
where the parameters $P_\mathrm{buy}$ and $P_\mathrm{sell}$ stand for the prices of electricity purchased from and sold to the grid, respectively, and the electricity trading costs are annualized using a scaling factor $F_\mathrm{anu}$. 
The parameters $C_\mathrm{pv}$ and  $C_\mathrm{bat}$ stand for the fixed costs (mainly installation costs) of PV and battery per unit installation amount per year, respectively.

The operation of the battery storage system is subject to the following  state equation: 
\begin{align}
& x_{\mathrm{soc}, k+1} =  x_{\mathrm{soc}, k} + E_\mathrm{chg}  u_{\mathrm{bin},k} - \dfrac{1}{E_\mathrm{dis}} u_{\mathrm{bout},k}, \quad \forall k  
\end{align}
where $x_{\mathrm{soc}, k}$ stands for the state of charge (SOC) of the battery storage at the time step $k$, and the parameters $E_\mathrm{chg}$ and $E_\mathrm{dis}$ for the efficiencies of charging and discharging of battery, respectively. 
In addition, the following constraints need to be addressed: 
\begin{align}
  & x_{\mathrm{soc},k} \le v_\mathrm{bat},  \quad \forall k,  \label{eq:soc_max} \\
  &  x_{\mathrm{soc}, N_\mathrm{t}+1 } = x_{\mathrm{soc}, 1} \label{eq:bondary_condition}
\end{align}
where \eqref{eq:soc_max} imposes an upper bound of SOC being subject to the long-term decision $v_\mathrm{bat}$, and \eqref{eq:bondary_condition}  is the boundary condition.  
Besides, the electricity demand should be equal to supply at each time step, and the following equation holds:
\begin{align} \label{eq:energy_balance}
  & u_{\mathrm{buy},k} - u_{\mathrm{sell},k}  -  u_{\mathrm{bin},k} + u_{\mathrm{bout},k} = D_k - \dfrac{S_k E_\mathrm{pv} v_\mathrm{pv}}{\color{red} G_\mathrm{stc}}, \, \forall k  
\end{align}
where $D_k$ stands for the electricity demand at the time step $k$ in {\color{red}$\si{kWh}$}, and $S_k$ for the solar radiation in {\color{red}$\si{kWh/m^2}$}. 
{\color{red}The parameter $G_\mathrm{stc}$ stands for the reference irradiance intensity of the Standard Test Conditions (STC) \cite{iec61724}, which is equal to $\SI{1000}{W/m^2}$,  
and $E_\mathrm{pv}$ stands for the performance ratio of PV system \cite{pearsall2017}, which} 
depends on the efficiencies of the PV panel and the power conditioning system and so on. 
An upper bound is imposed on the size of PV: 
\begin{align}
   v_\mathrm{pv} \le M_\mathrm{pv}.
\end{align}
This condition is needed to prevent the LP problem becomes unbounded when the PV price $C_\mathrm{pv}$ is lower than a certain value. 
Finally, the following condition forces all the decision variables to be non-negative:
\begin{align}
    v_\mathrm{pv}, \, v_\mathrm{bat}, \, u_{\mathrm{buy},k}, \, u_{\mathrm{sell},k},\, u_{\mathrm{bin},k}, \, u_{\mathrm{bout},k}, \, x_{\mathrm{soc},k}  \ge 0, \, \forall k. 
\end{align}

Note that the dimension of the above LP problem increases with the increase in the length of data analyzed. 
For an analysis over 1 year with a temporal resolution of 1 hour, the resultant LP problem becomes $2 + 8760 \times 5 = 43,802$ dimensional optimization problem, and may take hours or days to obtain the result by a general-purpose optimization solver.

\section{Screening Curve Method} \label{sec:scm}

In this section, we introduce the framework of SCM with a brief review of existing SCMs in \secref{sec:scm_thermal}. 
The proposed SCM for economic analysis of PV self-consumption is presented in  \secref{sec:scm_der}. 

\subsection{Review of Existing SCMs} \label{sec:scm_thermal}

\begin{figure}[!t]
 \centering
 \includegraphics[width=0.78\linewidth]{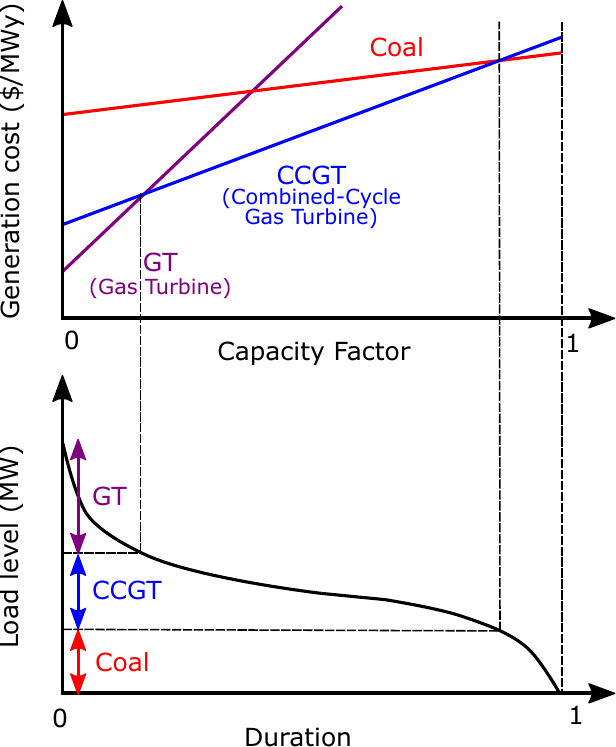}
 \caption{The classical SCM to determine the optimal thermal generation mix when no technical constraints are considered. } \label{fig:classical_scm}
\end{figure}

The classical SCM \cite{phillips69,lee78} requires two sets of input data: costs and a load curve for a target year. 
Cost data includes the annualized upfront investment cost $C_\mathrm{fix}$  per unit capacity (in \si{kW} or \si{MW}) and the variable production costs $C_\mathrm{var}$ per unit production (in \si{kWh} or \si{MWh}). 
The total annualized cost of a unit capacity of generation technology is represented as follows: 
\begin{align}
 c =C_\mathrm{fix}+C_\mathrm{var} T  \label{eq:cost_classical}
\end{align}
where $T$ stands for the firing hours of the considered thermal generation technology. 
The cost representation in \eqref{eq:cost_classical} is used to draw the so-called \emph{screening curve} for each generation technology as shown in the upper part of \figref{fig:classical_scm}, where the horizontal axis is taken as the capacity factor $f$ defined as the ratio of the firing hours $T$ to the total hours in a year ($f = T/\SI{8760}{h}$). 
\Figref{fig:classical_scm} illustrates examples of the screening curves for three candidate technologies: base-load (such as coal and nuclear), combined-cycle gas turbine (CCGT) and simple-cycle combustion turbine (GT).
Then, the intersections of these screening curves separate the regions of $f$ in which the different generation technologies become optimal. 
The least-cost generation mix for a given load curve can be obtained by combining them with the load-duration curve as shown in the lower part of  \figref{fig:classical_scm}, where the load-duration curve is obtained by sorting a chronological load curve in descending order of the load level. 
Since the load-duration curve maps a load level on the vertical axis to a capacity factor on the horizontal axis, and the screening curves map the capacity factor to the annual costs of generation alternatives, the optimal generation capacity mix can be read from the vertical axis of the lower graph as shown in \figref{fig:classical_scm}. 
This analysis gives the optimal solution to the least-cost problem with no technical constraints~\cite{stoft02}. 

\begin{figure}[!t]
    \begin{minipage}[b]{\linewidth}
    	  \centering
        \includegraphics[width=0.95\linewidth]{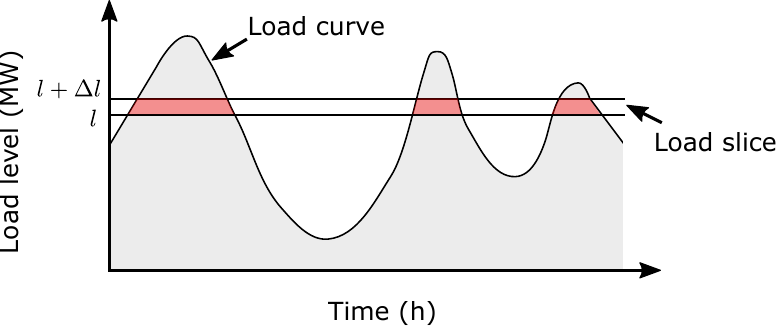}
        \subcaption{Definition of load slice } \label{fig:load_slice}
    \end{minipage}
   \begin{minipage}[b]{\linewidth}
        \centering
        \includegraphics[width=0.8\linewidth]{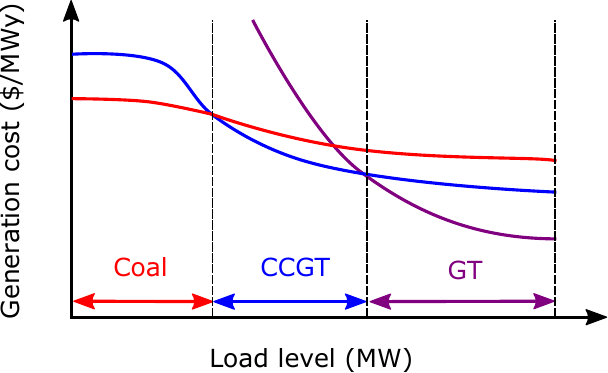}
        \subcaption{Example of screening curves (cost curves)} \label{fig:enhanced_scm} 
   \end{minipage}
   \caption{The framework of enhanced SCMs to estimate the optimal generation mix when technical constraints are considered. }
\end{figure}

In recent years, enhanced versions of SCM have been developed to estimate the optimal generation mix considering more detailed situations.
For this purpose, it was proposed in \cite{batlle13} to relate the load level directly to the generation cost. 
With this scheme, a chronological load curve is discretized into multiple load slices as shown in \figref{fig:load_slice}. 
For example, if the width of a load slice is set to be $\SI{0.1}{MW}$, then a load curve with a peak of $\SI{80}{MW}$ will be divided into $800$ load slices.
Then, the original least-cost problem will be decomposed into sub-problems corresponding to these slices, and a heuristic optimization is performed to determine which generation technology is used in each slice. 
This is done by calculating the annualized total cost to balance the demand in each load slice with a production by each generation technology to draw screening curves (cost curves) as shown in \figref{fig:enhanced_scm}. 
The estimated optimal generation mix can be derived by simply choosing the least-cost technology for each slice. 
With this approach, in addition to the fixed and variable costs in \eqref{eq:cost_classical}, a simple representation of the start-up costs can be obtained by counting the number of segments in each load slice and summing up the cost for each start up.  
In the example shown in \figref{fig:load_slice}, the costs for three re-starts will be added. 
Furthermore, the start-up can be separated into hot starts and cold starts depending on the shut-down duration before the restart, which is evaluated from the distance between adjacent segments \cite{batlle13,zhang15}. 
With the above framework of enhanced SCM, other technical details such as thermal cycling \cite{zhang15}, ancillary services \cite{zhang15:ancillary}, 
hydro scheduling \cite{staffell16}, existing capacities \cite{zhang17,guner18}, and planned outages \cite{zhang19} have been taken into account.

\subsection{Proposed SCM for PV and battery} \label{sec:scm_der}

Based on the framework of enhanced SCM introduced above, here we develop the proposed SCM for economic analysis of PV self-consumption. 

\subsubsection{Overview}

\begin{figure}[t]
 \centering
 \includegraphics[width=0.93\linewidth]{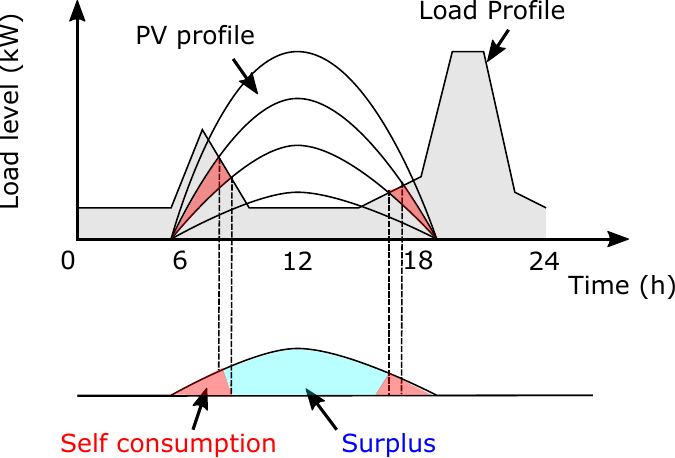}
 \caption{Time-varying scheme of load slicing in the proposed SCM} \label{fig:pv_load_slice}
\end{figure}

The existing SCMs have been developed mainly for thermal generation technologies and are not able to take into account intermittent nature of renewable energy sources. 
To consider the fact that the amount of electricity generated by PV depends on the solar irradiation at each time, we newly adopt a time-varying scheme for load slicing. 
Specifically, as shown in \figref{fig:pv_load_slice}, the load curve is discretized into multiple load slices using the generation profile of a unit capacity of PV, which corresponds to the width of the load slice. 
As shown in the lower part of the figure, the PV generation shaded in red will be directly consumed and the area shaded in blue becomes surplus electricity, which will be sold to the grid or charged to the battery to increase self-consumption. 
Note that the load demand during nighttime is not directly considered in this method, since there is no power generation by PV at night.  

With this modified approach of load slicing, the overall procedure of the analysis is followed from the framework of enhanced SCM and is given as follows:
\begin{enumerate}[(i)]
 \setlength{\parskip}{0cm} 
 \setlength{\itemsep}{0.1cm}  
 \item Discretize the chronological load curve into time-varying load slices by using the generation profile of a unit capacity of PV, denoted by $\Delta P$, corresponding to the width of a load slice.   
 \item Draw screening curves by calculating the annualized cost to satisfy the demand in each slice by each of the following three technology options: a) buying electricity from the grid, b) installing PV without battery, and c) installing PV with battery. 
 \item Estimate the optimal PV size by counting the number of slices where the option b) or c) becomes the least-cost technology. 
 \item Estimate the optimal battery size by summing up the required amount of battery calculated for each slice where the option c) becomes the least-cost technology.  
\end{enumerate}
The detailed calculations of the screening curves and required amount of battery are presented in the rest of this section. 
{\color{red}
Parameters and variables used for calculations are summarized in \tabref{tab:parameters_scm}. 
}

\begin{table}[t]%
\footnotesize
\caption{\color{red}Parameters and variables used for SCM}
\label{tab:parameters_scm}
\centering
\begin{minipage}{\linewidth}
\color{red}
\begin{tabular}{c|l }
\hline
Symbol & Meaning  \\
\hline
 $\Delta P$ & Unit capacity of PV corresponding to the width of a slice  \\
 $N_\mathrm{t}$ & Total number of time steps of analyzed data\\
 $N_\mathrm{d}$ & Total number of days of analyzed data \\ 
 $i$ & Index of load slices \\ 
 $k$ & Index of time steps in chronological order \\
 $j$ & Index of days in ascending order w.r.t. surplus electricity \\
 $q_{\mathrm{load},i,k}$ &  Electricity demand at the time step $k$ in the slice $i$ \\
 $q_{\mathrm{sur}, i,k}$ &  Surplus electricity at the time step $k$ in the slice $i$ \\
 $q_{\mathrm{chg},i,k}$ & Charged electricity at the time step $k$ in the slice $i$ \\
 $q_{\mathrm{bat},i}$ & Amount of battery required for the slice $i$ \\
 $Q_{\mathrm{bat},i,j}$ & Amount of battery required for the $j$-th day in the slice $i$ \\
 $\Delta Q_{\mathrm{bat},i,j}$ & Incremental battery amount for the $j$-th day in the slice $i$  \\
\hline
\end{tabular}
\end{minipage}
\end{table}

\subsubsection{Calculation of Screening Curves}

For the technology option a), the annualized cost $c_{\mathrm{grid},i}$ of purchasing electricity from the grid for the slice level $i$ can be calculated as follows: 
\begin{align} \label{eq:cost_grid}
 c_{\mathrm{grid}, i} = F_\mathrm{anu} \sum_{k=1}^{N_\mathrm{t}} P_\mathrm{buy}  q_{\mathrm{load}, i,k}
\end{align}
where $q_{\mathrm{load},i,k}$ stands for the amount of load demand of the slice $i$ at the time step $k$. 
The total amount of electricity $\sum_k q_{\mathrm{load}, i,k}$ corresponds to the area shaded in red in \figref{fig:pv_load_slice}. 
The parameters $P_\mathrm{buy}$ and $F_\mathrm{anu}$ are the same with those defined in \secref{sec:lp} and stand for the electricity retail price and the annualization factor, respectively.

For the technology option b), the annualized cost $c_{\mathrm{pv},i}$ of installing PV without battey for the slice $i$ is given as follows: 
\begin{align} 
 c_{\mathrm{pv},i} = C_\mathrm{pv} \Delta P - F_\mathrm{anu}  \sum_{k=1}^{N_\mathrm{t}} P_\mathrm{sell} q_{\mathrm{sur},i,k} \label{eq:cost_pv}
\end{align}
where $q_{\mathrm{sur},i,k}$ stands for the amount of surplus electricity for the slice $i$ at the time step $k$. 
The first term of the right-hand side of \eqref{eq:cost_pv} represents the investment cost for PV, and the second term the revenue by selling surplus electricity to the grid. 
The total amount of electricity $\sum_k q_{\mathrm{sur},i,k}$ corresponds to the area shaded in blue in \figref{fig:pv_load_slice}. 
The parameters $C_\mathrm{pv}$ and $P_\mathrm{sell}$ stand for the annualized fixed cost of PV and the price of selling electricity to the grid, respectively.

For the technology option c), the cost $c_{\mathrm{pv/bat},i}$ of installing PV and using battery for increasing self-consumption is given by 
\begin{align} \label{eq:cost_pv_battery}
 c_{\mathrm{pv/bat},i} = & \,
   C_\mathrm{pv}\Delta P  + C_\mathrm{bat} q_{\mathrm{bat},i}  \notag \\ 
   & - F_\mathrm{anu} \sum_{k=1}^{N_\mathrm{t}} P_{\mathrm{sell}} \left(  q_{\mathrm{sur},i,k} - q_{\mathrm{chg},i,k} \right)  \notag \\
  &  - F_\mathrm{anu} \sum_{k=1}^{N_\mathrm{t}} P_\mathrm{buy}  E_\mathrm{dis} E_\mathrm{chg}  q_{\mathrm{chg},i,k}   
\end{align}
where $q_{\mathrm{bat,}i}$ {\color{red}stands} for the required amount of battery for the slice $i$, and $q_{\mathrm{chg},i,k}$ stands for the amount of charged electricity for the slice $i$ at the time $k$. 
The procedure of estimation of these values will be presented later. 
The second term of the right-hand side of \eqref{eq:cost_pv_battery} represents the investment cost for battery. The third term represents the revenue by selling surplus electricity. The fourth term {\color{red}represents} the economic benefit due to the increase in self-consumption by charging surplus electricity and using it in e.g. nighttime. 
While the profile of charging is modeled in the proposed method, the discharging will not be explicitly considered. 

\subsubsection{Estimation of Battery Capacity and Charging Profile}

\begin{figure}[!t]
 \centering
 \subcaptionbox{Definition of the index $j$ \label{fig:def_j}}{
 \includegraphics[width=0.8\linewidth]{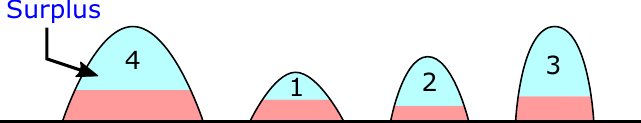} 
  }\\[5mm]
 \subcaptionbox{Charged electricity for the case of $j=1$ \label{fig:j=1}}{
   \includegraphics[width=0.8\linewidth]{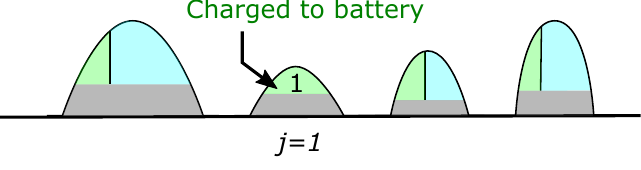} }\\[5mm]
 \subcaptionbox{Charged electricity for the cases of $j=2$ and $3$ \label{fig:j=2-3}}{
   \includegraphics[width=0.8\linewidth]{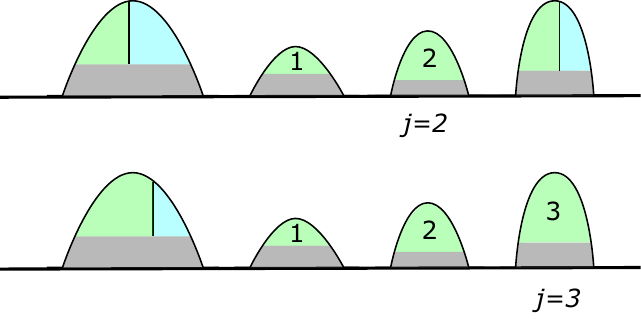} }
 \caption{Examples of charging profile of electricity for each value of $j=1, \, 2$, and $3$.} 
\end{figure}

To estimate the values of $q_{\mathrm{bat},i}$ and $q_{\mathrm{chg},i,k}$, we will consider the economic benefit of battery in a sequential manner. 
To this end, let us define an index $j$ that indicates the ascending order of the days with respect to the total amount of surplus PV generation in each day. 
That is, as shown in \figref{fig:def_j}, the day with the index $j=1$ has the smallest  amount of surplus electricity, and the day with $j=N_\mathrm{d}$ has the largest amount, where $N_\mathrm{d}$ stands for the number of days analyzed. 
This order may vary depending on the slice $i$, but we do not explicitly show it in the notation. 
Then, the amount of battery required to store all the surplus electricity of the $j$-th day can be given as 
\begin{align}
 Q_{\mathrm{bat}, i,j} := E_\mathrm{chg} \sum_{k \in \mathcal{K}_j}  q_{\mathrm{sur},i, k}     
\end{align}
where the symbol $\mathcal{K}_j$ stands for the set of indices of the time step $k$ belonging to the $j$-th day.
Note that this assumes that there is no discharging of battery during day time, otherwise the required amount of battery could be smaller.

%

Now the economic benefit of the battery can be sequentially estimated for each slice. 
First, consider to install a battery storage system of the amount of $Q_{\mathrm{bat},i,1}$ for the slice $\# i$. 
Then, the surplus electricity can be charged up to this amount every day (see \figref{fig:j=1}), and  
the economic benefit $B_1$ associated to this use of battery can be written as
\begin{align}
 B_1 =
  &  F_\mathrm{anu} N_\mathrm{d} P_\mathrm{buy} E_\mathrm{dis} Q_{\mathrm{bat},i, 1} \notag \\ 
  & - C_\mathrm{bat} Q_{\mathrm{bat},i, 1}  - F_\mathrm{anu} N_\mathrm{d}  P_\mathrm{sell} Q_{\mathrm{bat},i, 1}/E_\mathrm{chg}
\end{align}
where the first term represents the amount of payment of electricity that can be avoided, the second term the installation cost of the battery, and the third term the lost revenue that could have been earned by selling this electricity. 
Next, for $2  \le j \le N_\mathrm{d}$,  if the amount of battery is increased to $Q_{\mathrm{bat},i,j}$, the marginal economic benefit $B_j$ due to the incremental amount of battery defined by $\Delta Q_{\mathrm{bat},i,j} := Q_{\mathrm{bat},i,j} - Q_{\mathrm{bat},i,j-1}$ can be given by 
\begin{align}
 B_j = 
 &  F_\mathrm{anu} (N_\mathrm{d}-j+ 1)  P_\mathrm{buy} E_\mathrm{dis}\Delta Q_{\mathrm{bat},i, j}  \notag \\
 & - C_\mathrm{bat} \Delta Q_{\mathrm{bat},i, j}  \notag \\  &  - F_\mathrm{anu} (N_\mathrm{d} -j+ 1) P_\mathrm{sell} \Delta Q_{\mathrm{bat},i, j}/E_\mathrm{chg}.   \label{eq:b_j}
\end{align}
\Figref{fig:j=2-3} shows examples for the cases of $j = 2$ and $j=3$. 
Note that the economic benefit of the incremental battery capacity $\Delta Q_{\mathrm{bat},i,j}$ can occur only for $N_\mathrm{d} -j + 1$ days because the surplus electricity over the first to $(j-1)$-th days is already accounted in $B_1, \dots,\, B_{j-1}$. 
By rewriting \eref{eq:b_j}, we have 
\begin{align}
 B_j = 
 & \{ F_\mathrm{anu} (N_\mathrm{d}-j+ 1) (P_\mathrm{buy} E_\mathrm{dis}-P_\mathrm{sell} /E_\mathrm{chg})   \notag \\
 & - C_\mathrm{bat} \} \Delta Q_{\mathrm{bat},i, j}. 
\end{align}
Thus, the value of $ B_j/\Delta Q_{\mathrm{bat}i,j}$ decreases as the increase in $j$ when $P_\mathrm{buy} > P_\mathrm{sell} /(E_\mathrm{chg}E_\mathrm{dis})$ holds\footnote{Note that this condition holds for the setting in \tabref{tab:parameters} and is necessary for a self-consumption policy to work. Otherwise $B_j < 0$ for all $j = 1, \dots, N_\mathrm{d}$.}.  
%
%
%
Since the economic benefit of the incremental amount of battery $\Delta Q_{\mathrm{bat}, i, j}$ occurs when $B_j > 0$, the installation capacity of battery $q_{\mathrm{bat},i}$ in \eqref{eq:cost_pv_battery} can be given by 
\begin{align}
    q_{\mathrm{bat},i} := Q_{\mathrm{bat},i, J} = E_\mathrm{chg} \sum_{k \in \mathcal{K}_J} q_{\mathrm{sur}, i, k} \label{eq:battery_amount}
\end{align}
with
\begin{align}
    J := \max \{ j \in \mathbb{Z}_{[0, N_\mathrm{d}]} ~|~   B_j \ge 0  \} 
\end{align}
where $\mathbb{Z}_{[0, N_\mathrm{d}]}$ stands for the set of integers ranging from $0$ to $N_\mathrm{d}$, and for $j=0$, we define $B_0 := 0$ and $Q_{\mathrm{bat},i, 0} := 0$. 
From the observation that the value of $B_j/\Delta Q_{\mathrm{bat},i,j}$ decreases with $j$, the value of $J$ can be analytically described as 
\begin{align}
    J = \max\left( 0, \, \left\lfloor N_\mathrm{d}+1-\dfrac{C_\mathrm{bat} E_\mathrm{chg} }{F_\mathrm{anu}(P_\mathrm{buy}E_\mathrm{dis} E_\mathrm{chg} -P_\mathrm{sell} )} \right\rfloor \right) \label{eq:J_analytic}
\end{align}
where $\lfloor \cdot \rfloor$ stands for the floor function. 

Given $J$ and $q_{\mathrm{bat},i}$ above, we can arbitrarily determine  the charging profile $\{  q_{\mathrm{chg}, i, k} \}_{k=1}^{N_\mathrm{t}}$ used in \eqref{eq:cost_pv_battery}  as long as the total charged amount of electricity at each day does not exceed the upper limit given by $q_{\mathrm{bat},i}/E_\mathrm{chg}$. 
Furthermore, by defining the total amount of electricity $Q_{\mathrm{chg}, i}$ charged to the battery as 
\begin{align}
 Q_{\mathrm{chg},i} =  \sum_{j=1}^{J} \sum_{k \in \mathcal{K}_j } q_{\mathrm{sur},i, k} + (N_\mathrm{d} -J) \sum_{k \in \mathcal{K}_J }  q_{\mathrm{sur},i, k},  \label{eq:Q_charge}
\end{align}
the cost \eqref{eq:cost_pv_battery} can be rewritten as 
\begin{align}
 c_{\mathrm{pv/bat},i}  = & \,
   C_\mathrm{pv}\Delta P  + C_\mathrm{bat} E_\mathrm{chg} \sum_{k \in \mathcal{K}_J} q_{\mathrm{sur}, i, k} \notag \\ 
   & - F_\mathrm{anu} \sum_{k=1}^{N_\mathrm{t}} P_{\mathrm{sell}} q_{\mathrm{sur},i,k} \notag \\
  &  - F_\mathrm{anu} (P_\mathrm{buy}  E_\mathrm{dis} E_\mathrm{chg}-P_\mathrm{sell})  Q_{\mathrm{chg},i}. 
  \label{eq:cost_battery_final}
\end{align}
Thus, the screening curve of $c_{\mathrm{bv/bat},i}$ can be directly calculated from the profile $\{q_{\mathrm{sur}, i,k} \}_{k=1}^{N_\mathrm{t}}$ by using \eqref{eq:J_analytic}, \eqref{eq:Q_charge}, and \eqref{eq:cost_battery_final} without calculating the values of $B_j$.

\section{Verification Study} \label{sec:verification}

This section examines the estimation accuracy of the proposed SCM by comparing the results with the LP problem presented in \secref{sec:lp}. 
The python source code for the implementation of the proposed SCM is available upon request. 

\subsection{Base-Case Analysis}

\begin{figure}[!t]
 \centering
 \subcaptionbox{PV data}{
 \includegraphics[width=0.97\linewidth]{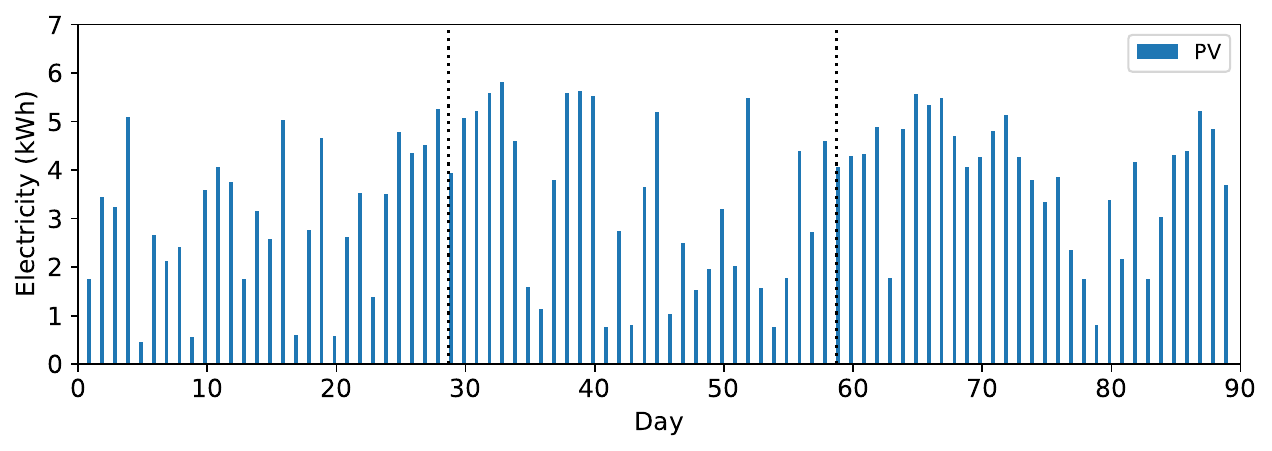}
  }\\[2mm]
 \subcaptionbox{Demand data}{
   \includegraphics[width=0.97\linewidth]{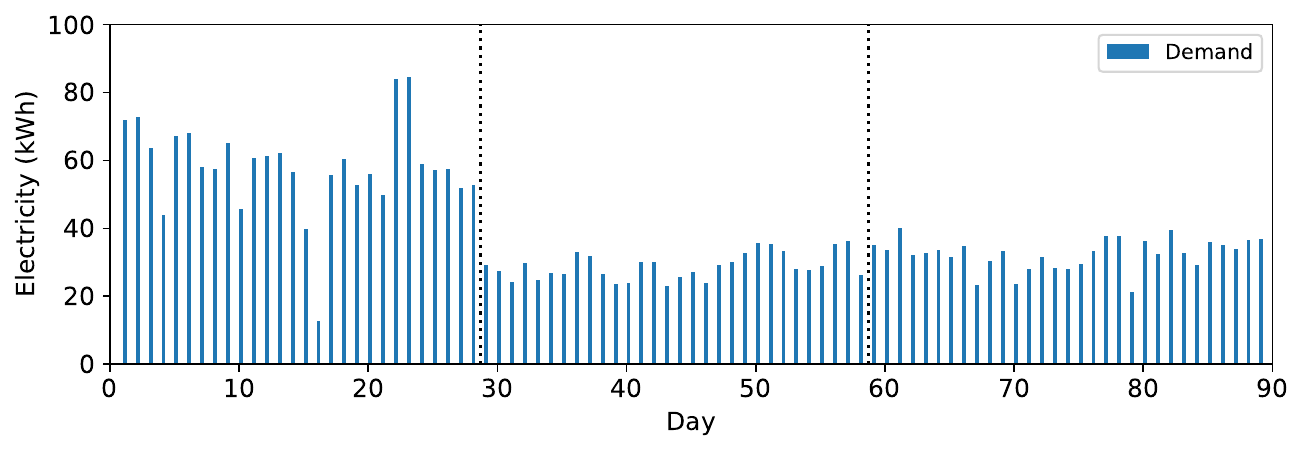} }
 \caption{Daily profiles of the input PV and demand data {\color{red}for February (1--28th days), June (29--58th days), and August (59--89th days)}.} \label{fig:input_data}
\end{figure}

The base-case analysis is performed with the setting of parameters shown in \tabref{tab:parameters}. 
A synthetic hourly load profile was generated from real measurement data of a household in Japan \cite{database} by concatenating data for February (Winter), June (Rainy season), and August (Summer) to construct a three-month data including different seasons in a year. 
A generation profile for a unit capacity of PV was generated by using solar irradiation data of METPV20 \cite{metpv20} for corresponding days and the location of the load profile. 
The daily profiles of these input data are shown in \figref{fig:input_data}. 
It can be seen that the solar irradiation is generally low in winter, and there are many days with low solar irradiation in the rainy season. 
On the other hand, demand is higher during winter than other seasons. 
{\color{red}Although we use demand and solar irradiation data from Japan, it contains wide spectrum of situation  of weather and seasons, and we aim to verify that the proposed method works also for other regions or countries.}

\begin{figure}[!t]
 \centering
 \subcaptionbox{Cost curves \label{fig:cost_curves}}{
 \includegraphics[width=0.85\linewidth]{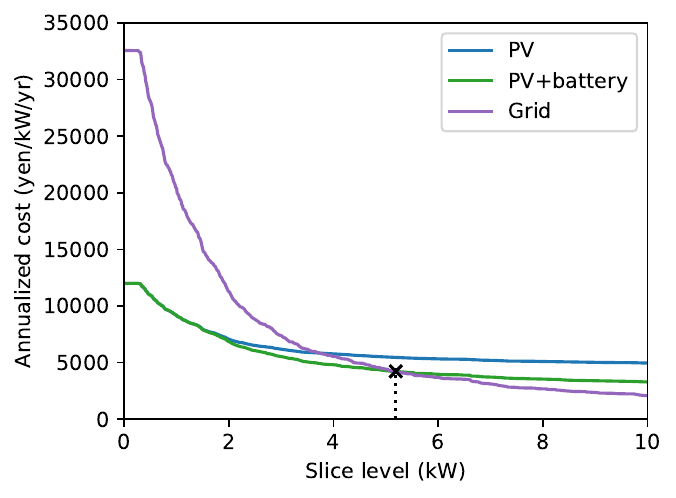}
 }\\[2mm]
 \subcaptionbox{\color{red}Cumulative amount of battery \label{fig:cummlative battery}}{
 \hspace{3mm}\includegraphics[width=0.8\linewidth]{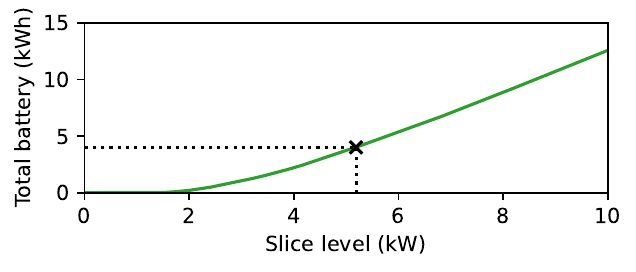}
 } 
 \caption{Base-case result of the proposed SCM} 
\end{figure}

The screening curves (cost curves) obtained by the proposed SCM is shown in {\color{red}\figref{fig:cost_curves}}. 
The \emph{green} and \emph{blue} lines in the figure show the cost curves for using PV with and without battery storage calculated by \eqref{eq:cost_battery_final} and \eqref{eq:cost_pv}, respectively, and the \emph{purple} line the cost curve for buying electricity from the grid calculated by \eqref{eq:cost_grid}. 
The cost curve of ``PV+battery" (green line) takes least value from $l= 0$ to $\SI{5.19}{kW}$ and has an intersection with the cost curve of ``Grid" (purple line) at the point shown by the cross symbol ($\times$) in the figure.  
Thus, the installation capacity of PV can be estimated as $\SI{5.19}{kW}$. 
{\color{red}
The amount of battery is calculated for each slice by using \eqref{eq:battery_amount}. 
\Figref{fig:cummlative battery} shows the cumulative amount of battery $\sum_i q_{\mathrm{bat},i}$ until corresponding slice level.
The total amount of battery can be derived by considering the level of $l=\SI{5.19}{kW}$, and is given as $\SI{4.01}{kWh}$.}
The optimal PV and battery capacities obtained by solving the LP problem are $\SI{5.15}{kW}$ and $\SI{4.06}{kWh}$, respectively, and the proposed SCM approximates these optimal values appropriately. 

{\color{red}
In \figref{fig:cost_curves}, the screening curves of PV and PV+battery start at the same point (same annualized cost at $\SI{0}{kW}$ slice level). 
This is because at $\SI{0}{kW}$ slice level, the surplus electricity $q_{\mathrm{sur}, 0, k}$ is zero for all $k = 1, \dots N_\mathrm{t}$ (see \figref{fig:pv_load_slice}), and thus only the first terms of Eqs.\,\eqref{eq:cost_pv} and \eqref{eq:cost_battery_final} become nonzero. 
When the screening curves are plotted in \figref{fig:cost_curves}, the annualized costs are scaled by dividing by $\Delta P$, and the above two screening curves start with $C_\mathrm{pv}=\SI{12000}{yen/(kW\cdot yr)}$. 
Further, the option of PV+battery is always cheaper than the option of just using PV, since the amount of battery is determined such that the economic benefit $B_j$ becomes non-negative.
The difference between these two screening curves represents the amount of economic benefit owing to the battery.  
}

\begin{figure}[!t]
 \centering
 \subcaptionbox{Electricity sold to the grid}{
 \includegraphics[width=0.97\linewidth]{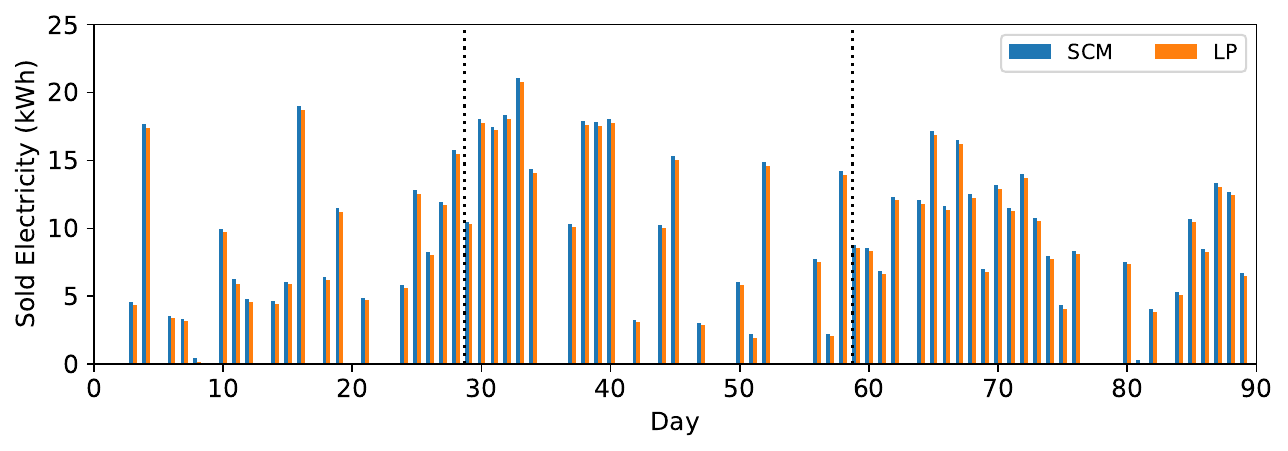}
  }\\[2mm]
 \subcaptionbox{Electricity charged to the battery storage}{
   \includegraphics[width=0.97\linewidth]{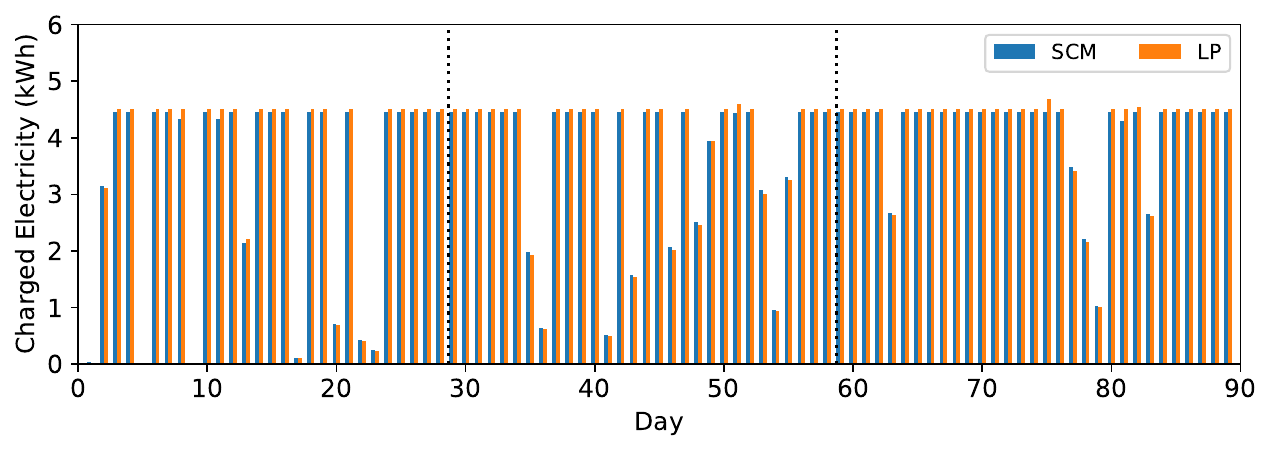} }
 \caption{Amounts of electricity soled and charged for each day.} \label{fig:profile}
\end{figure}

\Figref{fig:profile} shows the daily profiles of (a) electricity sold to the grid and (b) electricity charged to the battery.  
The profiles obtained by the SCM (shown by \emph{blue}) capture the optimal profiles by the LP problem (shown by \emph{orange}). 
It can be seen that, in many days, the amount of charged electricity takes its maximum value ($\SI{4.06}{kWh}/E_\mathrm{chg}=\SI{4.51}{kWh}$) meaning that the battery is fully charged. 
Although the SCM does not consider the profile of discharging, the ratio of the amount of sold electricity and the amount of charged electricity is properly reproduced throughout the analysis period, including the days when the amount of charge is low as typically seen in winter or rainy season.

\subsection{Estimation Accuracy under Various Settings}

Next, the estimation accuracy of the proposed SCM is examined under various settings of parameters. 
To reduce the computational time for obtaining the optimal solutions of the LP problem, the load demand and PV generation profiles were shortened by taking the first ten days of each month in the three-month data used above to construct a one-month long data.
Among the parameters shown in \tabref{tab:parameters}, the four parameters $C_\mathrm{pv}$, $C_\mathrm{bat}$, $P_\mathrm{buy}$, and $P_\mathrm{sell}$, which are directly relevant to the regulatory discussion of the PV self-consumption, are changed from the setting of the base-case analysis.  

\begin{figure}[!t]
 \centering
 \subcaptionbox{Changes in $C_\mathrm{pv}$   \label{fig:diff_pv}}{
   \includegraphics[width=0.87\linewidth]{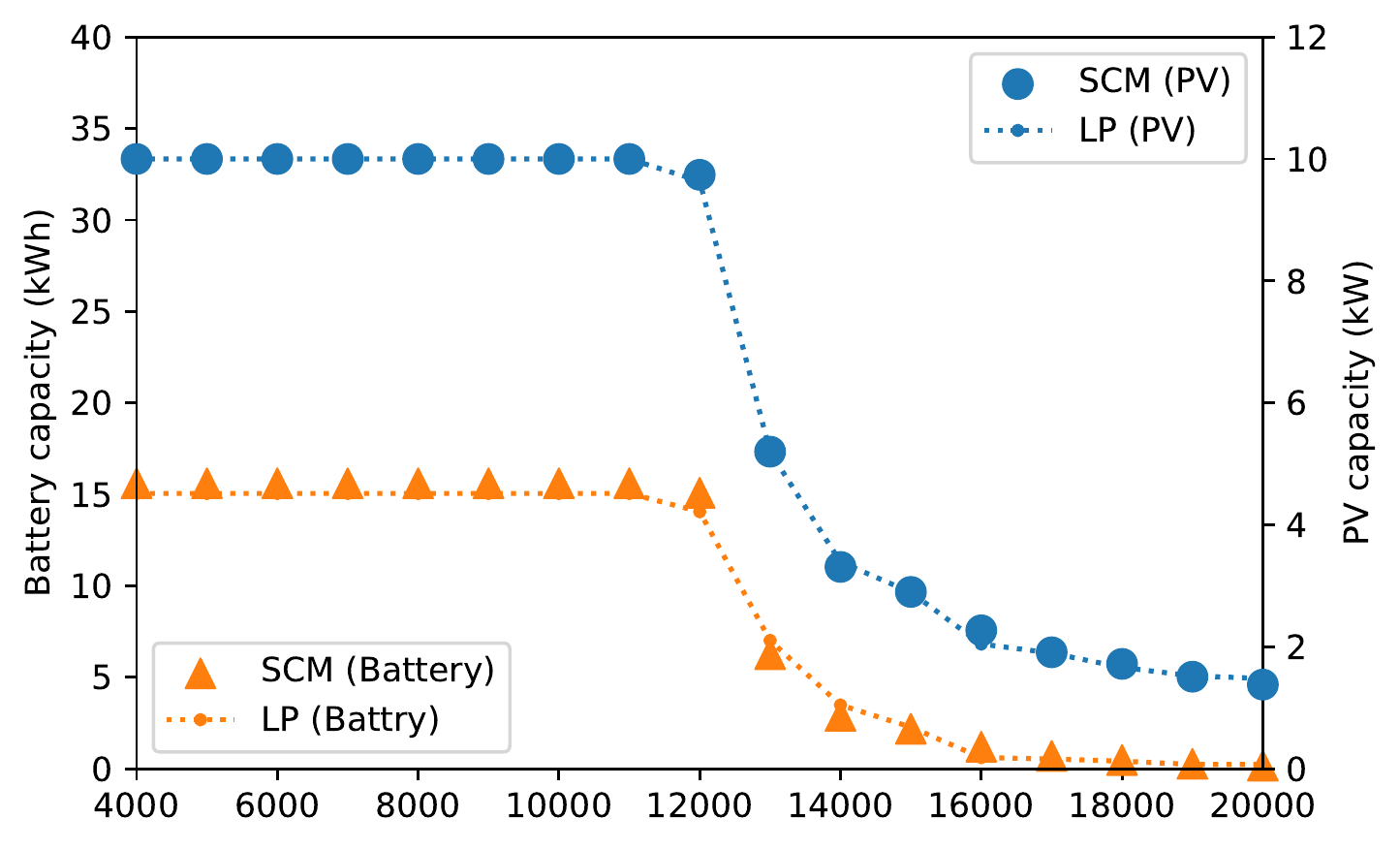} }\\[2mm]
 \subcaptionbox{Changes in $C_\mathrm{bat}$ \label{fig:diff_bat}}{
  \includegraphics[width=0.87\linewidth]{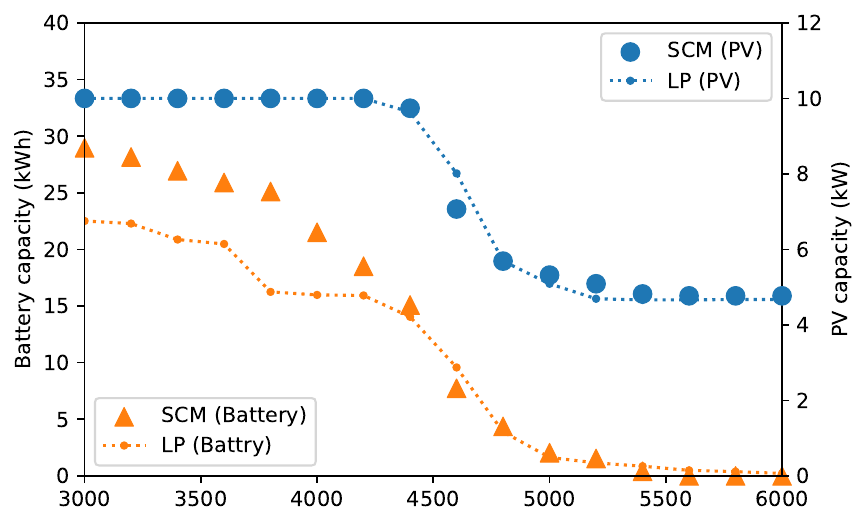} }
 \caption{Results with various setting of PV and battery prices. } \label{fig:diff_fixed_costs}
\end{figure}

\Figref{fig:diff_fixed_costs} shows the comparison between the estimated and optimal solutions under varying prices of PV and battery storage systems. 
\Figref{fig:diff_pv} shows the results of changing the setting of the parameter $C_\mathrm{pv}$. 
The \emph{blue circles} show the PV capacities estimated by the proposed SCM, and they are close to the results of the LP problem shown by the dotted line. 
Similarly, the estimated battery capacities shown by the \emph{orange triangles} are close to the optimal capacities over the entire range of $C_\mathrm{pv}$. 
%
Next, \figref{fig:diff_bat} shows the result of changing the setting of the battery fixed cost $C_\mathrm{bat}$.  
While the estimated capacity of PV is close to the optimal value over the entire range of $C_\mathrm{bat}$, the installation capacity of battery is overestimated for the region where $C_\mathrm{bat}$ is lower than $\SI{4000}{yen/kWh/yr}$.
A plausible reason for this is that, while the proposed SCM does not take into account the profile of discharging, the battery is not sufficiently discharged each day, and there is no room for charging as much as calculated by the SCM.
{\color{red}Thus, the solution derived by the SCM becomes suboptimal for the household.}  
Nevertheless, the proposed SCM capture the overall trend of the changes in the optimal battery size. 

\begin{figure}[!t]
 \centering
 \subcaptionbox{Change in $P_\mathrm{buy}$   \label{fig:diff_pbuy}}{
   \includegraphics[width=0.87\linewidth]{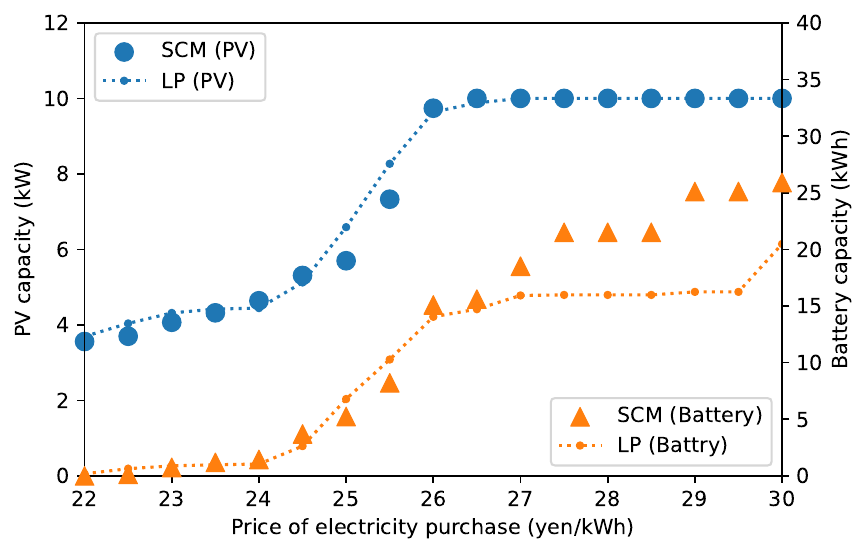} }\\[2mm]
 \subcaptionbox{Change in $P_\mathrm{sell}$ \label{fig:diff_psell}}{
  \includegraphics[width=0.87\linewidth]{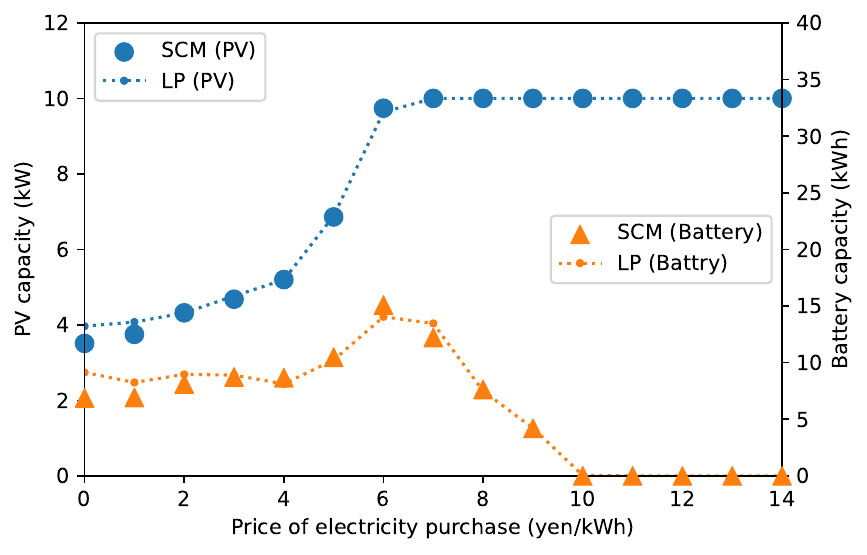} }
 \caption{Results with various setting of electricity prices.} \label{fig:diff_prices}
\end{figure}

\Figref{fig:diff_prices} shows the comparison of the estimated and optimal capacities under various setting of electricity prices. 
\Figref{fig:diff_pbuy} shows the results where the retail price $P_\mathrm{buy}$ is changed. 
It can be seen that the proposed SCM captures the overall trend of the changes in the optimal values of PV and battery, while the estimation accuracy of the battery capacity deteriorates in the region where the price $P_\mathrm{buy}$ is higher than $\SI{27}{yen/kWh}$. 
As in the case of \figref{fig:diff_bat}, it starts to overestimate the economic benefit of the battery when the battery capacity becomes greater than $\SI{15}{kWh}$. 
Finally, \figref{fig:diff_psell} shows the results where the price $P_\mathrm{sell}$ is changed. 
The estimated and optimal values are closed to each other for both PV and battery capacities. 

{\color{red}
To summarize, the proposed SCM can estimate the optimal installation capacities of PV and battery for a wide range of parameter settings of $C_\mathrm{pv}$, $C_\mathrm{bat}$, $P_\mathrm{buy}$, and $P_\mathrm{sell}$, except for a region where the battery size becomes larger than a certain level. 
While it was assumed that there is no discharging of battery during day time, it did not largely affect the estimation accuracy. 
However, it should be noted that this assumption may become a limitation when the prices of buying and selling electricity do not remain the same during the day and at night.  
}

\section{Discussion} \label{sec:discussion}

In this section, we sort out advantages and disadvantages of the proposed SCM and discuss its effectiveness.

\subsection{Economic Analysis Using Screening Curves}

Here we illustrate how the cost curves obtained by the proposed SCM can be used for understanding and predicting the changes in the optimal capacities of PV and battery storage systems. 
Although similar examples have been presented in \cite{hoshino21}, we provide recalculated results using the  SCM  proposed in \secref{sec:scm} to show its effectiveness in a self-consistent way. 

\begin{figure}[!t]
 \centering
   \includegraphics[width=0.85\linewidth]{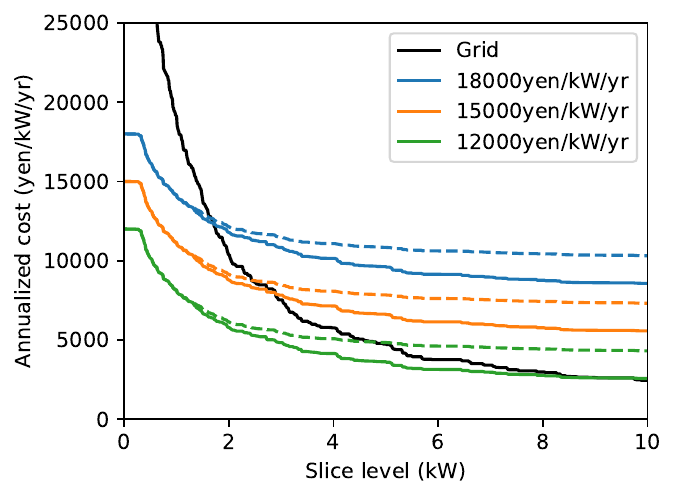}  \caption{Cost curves with different values of $C_\mathrm{pv}$   \label{fig:cost_curves_diff_pv}} 
\end{figure}

\Figref{fig:cost_curves_diff_pv} shows the screening curves calculated under the three settings of the price of PV: $C_\mathrm{pv} = \SI{18000}{yen/(kW\cdot yr)}$, $\SI{15000}{yen/(kW\cdot yr)}$, and $\SI{12000}{yen/(kW\cdot yr)}$. 
Since the cost curve of $c_{\mathrm{grid},i}$ is independent of $C_\mathrm{pv}$, it is common for all the above settings and shown by the \emph{black} line. 
The cost curves of $c_{\mathrm{pv},i}$ and $c_{\mathrm{pv/bat},i}$ are shown by \emph{broken} and \emph{solid} lines, respectively, and they move down as $C_\mathrm{pv}$ increases without changing their shapes. 
This is because the parameter $C_\mathrm{pv}$ only affects the first term of \eqref{eq:cost_pv} or  \eqref{eq:cost_battery_final}, and they do not depend on the slice level $i$. 
It can be seen that the intersection of the cost curves of $c_{\mathrm{pv},i}$ and $c_{\mathrm{grid},i}$ moves to the right as $C_\mathrm{pv}$ decreases, implying the increase in the installation capacities of PV and battery.   
Especially, when $C_\mathrm{pv}$ takes a value near $\SI{12000}{yen/kW/yr}$, the slope of the cost curve $c_{\mathrm{grid},i}$ at the intersection becomes small, and thus the optimal capacities of PV and battery drastically  changes as also can be seen in \figref{fig:diff_pv}.

\begin{figure}[!t]
 \centering
  \includegraphics[width=0.85\linewidth]{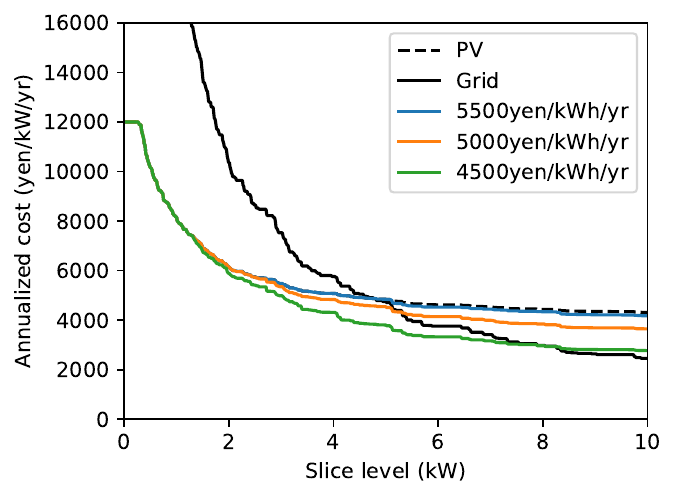} \caption{Cost curves with different values of $C_\mathrm{bat}$ \label{fig:cost_curves_diff_bat}} 
\end{figure}

Next, \figref{fig:cost_curves_diff_bat} shows the cost curves calculated under the settings of $C_\mathrm{bat}= \SI{5500}{yen/(kWh\cdot yr)}$, $\SI{5000}{yen/(kWh \cdot yr)}$, and $\SI{4500}{yen/(kWh\cdot yr)}$. 
The cost curves of $c_{\mathrm{grid},i}$ and $c_{\mathrm{pv},i}$ are invariant and shown by \emph{solid} and \emph{broken black} lines, respectively. 
The cost curve of $c_{\mathrm{pv/bat},i}$ goes down as $C_\mathrm{bat}$ decreases. 
The difference between $c_{\mathrm{pv},i}$ and $c_{\mathrm{pv/bat},i}$ describes the economic benefit of the self-consumption owing to battery storage, and it can be explained from this figure that the capacity of the battery storage begins to increase from zero at a point around  $\SI{5000}{yen/(kWh\cdot yr)}$ as can be seen in \figref{fig:diff_bat}.

As shown in these examples, the effect of changes in a parameter on each cost curve can be easily predicted from their definitions in \eqref{eq:cost_grid}, \eqref{eq:cost_pv}, and \eqref{eq:cost_battery_final}, 
and similar implications can also be obtained when other parameters change. 
By considering the behavior of each cost curve separately, the changes in the resultant optimal capacities in PV and battery storage systems can be understood in an intuitive and consistent way.

\subsection{Computational Time}

\begin{table}[!t]
 \centering
 \caption{Computational times}
 \label{tab:time}
 \begin{tabular}{ c | c c }
 Data length & Comp. Time (SCM) &  Comp. Time (LP) \\
 \hline 
 1 months & \SI{1.29}{s} & \SI{57.73}{s} \\ 
 2 months & \SI{2.46}{s} & \SI{276.15}{s} \\  
 3 months & \SI{2.93}{s} & \SI{784.18}{s} \\  
 4 months & \SI{3.49}{s} & \SI{1439.31}{s} \\  
 5 months & \SI{3.96}{s} & \SI{3289.80}{s} \\  
 6 months & \SI{4.56}{s} & \SI{6089.13}{s} \\   
\end{tabular}
\end{table}

\begin{figure}[!t]
 \centering
 \includegraphics[width=0.75\linewidth]{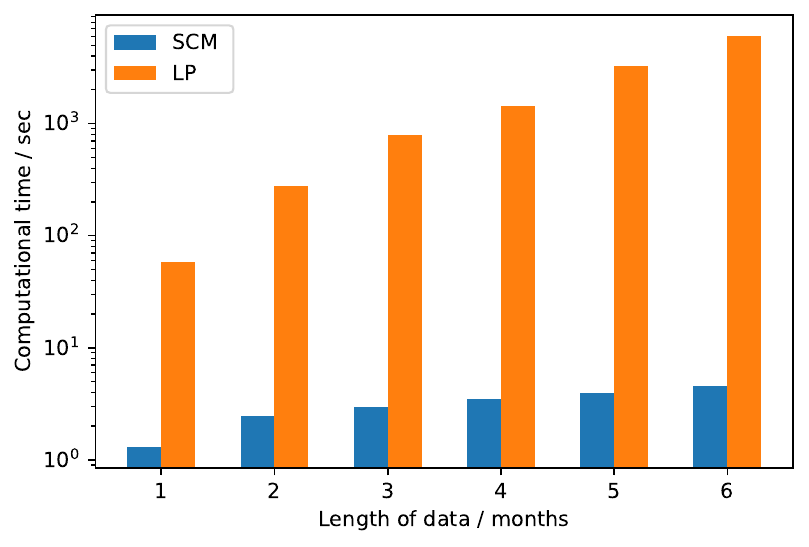}
 \caption{Computational times with SCM and LP} \label{fig:time}
\end{figure}

Since the computations of cost curves in the proposed SCM only requires numerical operations on arrays, it is faster than solving a least-cost problem using a general-purpose optimization solver.
\tabref{tab:time} and \figref{fig:time} show the comparison of the  computational times between the proposed SCM and the LP problem. 
These results have been obtained in simulations with a Core i9-10850K based desktop computer. 
The proposed SCM is much faster than solving the LP problem. 
The LP problem has been solved by using an interior-point method provided by \texttt{linprog} in the python package SciPy, and the computational time could be reduced by using much more powerful commercial solvers. 
However, when the length of data is increased, the rapid increase in the computational time cannot be avoided due to the increase in the dimension of the problem. 
The proposed SCM enables quick estimation of the optimization results without any special commercial solver and its computational time increases only linearly to the length of the data. 
The short computational time is useful to perform sensitivity analysis or other studies that require multiple simulations with different settings of parameters or input data.

\subsection{Advantages and Disadvantages}

The proposed SCM estimates the economic benefit of the photovoltaic self-consumption using battery storage.   
It does not consider other types of battery usage, such as energy arbitrage in response to variations in electricity prices \cite{telaretti16,zurfi17}  or provision of ancillary services for voltage regulation or congestion relief of distribution grid \cite{maeyaert20}. 
There are also limitations to the proposed SCM.
For example, it cannot describe discrete installation capacities, existing capacities, or a common three-part tariff, while some of these issues including ancillary services and existing capacities may be resolved by further modifications as performed with the SCM for thermal generation \cite{zhang15,zhang17}. 

One advantage of the proposed SCM is that it can provide intuitive plots of cost curves to understand the reasons behind the results of optimal installation capacities.
In addition, the computational time of the SCM is short (few seconds) compared to optimization models (usually more than hours). 
These facts are useful to explore cost sensitivity or other studies that require multiple simulations.
Thus, the proposed SCM is more useful for researchers or policy makers studying incentive policy for supporting PV or tariff regulations, who want to draw policy implications from an intuitive and quick trend analysis, rather than residential or commercial consumers who want to derive an optimal energy systems design with detailed specifications taken into account. 
Nevertheless, the SCM can be used to derive an initial estimate prior to the detailed design or to screen out cases where there is no economic benefit of installing  PV or battery storage systems.

\section{Conclusion} \label{sec:conclusion}

This paper developed a method of economic analysis of PV self-consumption with and without battery storage.  The development is based on a generalization of the framework of Screening Curve Method (SCM), which has been utilized for generation expansion problem of a bulk power system. 
The proposed SCM enables quick and intuitive estimation of the installation capacities of PV and battery storage systems. 
The estimation accuracy has been validated through numerical simulations, and high accuracy has been achieved when the battery installation capacity is not large. 
The method is useful to explore cost sensitivity or other studies that require multiple simulations.

Future directions of this study include further improvements of the method to consider, e.g., the usage of battery storage for ancillary services or the effect of existing capacities. 
Another one is to perform practical economic analyses by using the proposed SCM.

\section*{Acknowledgments}

The authors wish to thank K.~Okada and K.~Furusawa (Central Research Institute of Electric Power Industry; CRIEPI) for feedback during the course of this work.

\bibliography{main}

\begin{thebibliography}{10}
\providecommand{\url}[1]{#1}
\csname url@samestyle\endcsname
\providecommand{\newblock}{\relax}
\providecommand{\bibinfo}[2]{#2}
\providecommand{\BIBentrySTDinterwordspacing}{\spaceskip=0pt\relax}
\providecommand{\BIBentryALTinterwordstretchfactor}{4}
\providecommand{\BIBentryALTinterwordspacing}{\spaceskip=\fontdimen2\font plus
\BIBentryALTinterwordstretchfactor\fontdimen3\font minus
  \fontdimen4\font\relax}
\providecommand{\BIBforeignlanguage}[2]{{%
\expandafter\ifx\csname l@#1\endcsname\relax
\typeout{** WARNING: IEEEtran.bst: No hyphenation pattern has been}%
\typeout{** loaded for the language `#1'. Using the pattern for}%
\typeout{** the default language instead.}%
\else
\language=\csname l@#1\endcsname
\fi
#2}}
\providecommand{\BIBdecl}{\relax}
\BIBdecl

\bibitem{timilsina12}
G.~R. Timilsina, L.~Kurdgelashvili, and P.~A. Narbel, ``Solar energy: Markets,
  economics and policies,'' \emph{Renewable and Sustainable Energy Reviews},
  vol.~16, no.~1, pp. 449--465, 2012.

\bibitem{poruschi18}
L.~Poruschi, C.~L. Ambrey, and J.~C. Smart, ``Revisiting feed-in tariffs in
  australia: A review,'' \emph{Renewable and Sustainable Energy Reviews},
  vol.~82, pp. 260--270, 2018.

\bibitem{iea-pvps22}
{International Energy Agency}, ``Trends in photovoltaic applications 2022,''
  \url{https://iea-pvps.org/trends_reports/trends-2022/} (accessed 6/11/2023).

\bibitem{luthander15}
R.~Luthander, J.~Widén, D.~Nilsson, and J.~Palm, ``Photovoltaic
  self-consumption in buildings: A review,'' \emph{Applied Energy}, vol. 142,
  pp. 80--94, 2015.

\bibitem{ordonez22}
A.~{Ord\'{o}\~{n}ez}, E.~S\'{a}nchez, L.~Rozas, R.~Garc\'{i}a, and
  J.~Parra-Dom\'{i}nguez, ``Net-metering and net-billing in photovoltaic
  self-consumption: The cases of ecuador and spain,'' \emph{Sustainable Energy
  Technologies and Assessments}, vol.~53, p. 102434, 2022.

\bibitem{felder14}
F.~A. Felder and R.~Athawale, ``The life and death of the utility death
  spiral,'' \emph{The Electricity Journal}, vol.~27, no.~6, pp. 9--16, 2014.

\bibitem{schittekatte18}
T.~Schittekatte, I.~Momber, and L.~Meeus, ``Future-proof tariff design:
  Recovering sunk grid costs in a world where consumers are pushing back,''
  \emph{Energy Economics}, vol.~70, pp. 484--498, 2018.

\bibitem{ines20}
C.~In\^{e}s, P.~L. Guilherme, M.-G. Esther, G.~Swantje, H.~Stephen, and
  H.~Lars, ``Regulatory challenges and opportunities for collective renewable
  energy prosumers in the eu,'' \emph{Energy Policy}, vol. 138, p. 111212,
  2020.

\bibitem{dadamo22}
I.~D'Adamo, M.~Gastaldi, and P.~Morone, ``Solar collective self-consumption:
  Economic analysis of a policy mix,'' \emph{Ecological Economics}, vol. 199,
  p. 107480, 2022.

\bibitem{han22}
X.~Han, J.~Garrison, and G.~Hug, ``Techno-economic analysis of pv-battery
  systems in switzerland,'' \emph{Renewable and Sustainable Energy Reviews},
  vol. 158, p. 112028, 2022.

\bibitem{zhang17:pv}
Y.~Zhang, P.~E. Campana, A.~Lundblad, and J.~Yan, ``Comparative study of
  hydrogen storage and battery storage in grid connected photovoltaic system:
  Storage sizing and rule-based operation,'' \emph{Applied Energy}, vol. 201,
  pp. 397--411, 2017.

\bibitem{nyholm16}
E.~Nyholm, J.~Goop, M.~Odenberger, and F.~Johnsson, ``Solar
  photovoltaic-battery systems in swedish households – self-consumption and
  self-sufficiency,'' \emph{Applied Energy}, vol. 183, pp. 148--159, 2016.

\bibitem{schopfer18}
S.~Schopfer, V.~Tiefenbeck, and T.~Staake, ``Economic assessment of
  photovoltaic battery systems based on household load profiles,''
  \emph{Applied Energy}, vol. 223, pp. 229--248, 2018.

\bibitem{cervantes18}
J.~Cervantes and F.~Choobineh, ``Optimal sizing of a nonutility-scale solar
  power system and its battery storage,'' \emph{Applied Energy}, vol. 216, pp.
  105--115, 2018.

\bibitem{phillips69}
D.~Phillips, F.~P. Jenkin, J.~A.~T. Pritchard, and K.~Rybicki, ``A mathematical
  model for determining generating plant mix,'' in \emph{3rd IEEE Power Systems
  Computations Conference}, 1969.

\bibitem{lee78}
S.~T. Lee and C.~Dechamps, ``Mathematical model for economic evaluation of
  tidal power in the bay of fundy,'' \emph{IEEE Transactions on Power Apparatus
  and Systems}, vol. PAS-97, no.~5, pp. 1769--1778, 1978.

\bibitem{batlle13}
C.~Batlle and P.~Rodilla, ``An enhanced screening curves method for considering
  thermal cycling operation costs in generation expansion planning,''
  \emph{IEEE Transactions on Power Systems}, vol.~28, no.~4, pp. 3683--3691,
  2013.

\bibitem{zhang15}
T.~Zhang, R.~Baldick, and T.~Deetjen, ``Optimized generation capacity expansion
  using a further improved screening curve method,'' \emph{Electric Power
  Systems Research}, vol. 124, pp. 47--54, 2015.

\bibitem{zhang15:ancillary}
T.~Zhang and R.~Baldick, ``Consideration of ancillary services in screening
  curve method,'' in \emph{2015 IEEE Power \& Energy Society General Meeting},
  2015, pp. 1--5.

\bibitem{staffell16}
I.~Staffell and R.~Green, ``Is there still merit in the merit order stack? the
  impact of dynamic constraints on optimal plant mix,'' \emph{IEEE Transactions
  on Power Systems}, vol.~31, no.~1, pp. 43--53, 2016.

\bibitem{zhang17}
T.~Zhang and R.~Baldick, ``Consideration of existing capacity in screening
  curve method,'' \emph{IEEE Transactions on Power Systems}, vol.~32, no.~4,
  pp. 3038--3048, 2017.

\bibitem{guner18}
Y.~E. G\"{u}ner, ``The improved screening curve method regarding existing
  units,'' \emph{European Journal of Operational Research}, vol. 264, no.~1,
  pp. 310--326, 2018.

\bibitem{zhang19}
T.~Zhang and R.~Baldick, ``Generation planning using a modified screening curve
  method to account for planned outage,'' in \emph{2019 IEEE Power \& Energy
  Society General Meeting (PESGM)}, 2019, pp. 1--5.

\bibitem{stoft02}
S.~Stoft, \emph{Power System Economics: Designing Markets for
  Electricity}.\hskip 1em plus 0.5em minus 0.4em\relax Wiley-IEEE Press, 2002.

\bibitem{priya14}
G.~{Krishna Priya} and S.~Bandyopadhyay, ``Screening curve method for optimum
  source sizing to satisfy time varying demand,'' \emph{Computer Aided Chemical
  Engineering}, vol.~33, pp. 1573--1578, 2014.

\bibitem{ru14}
Y.~Ru, J.~Kleissl, and S.~Martinez, ``Exact sizing of battery capacity for
  photovoltaic systems,'' \emph{European Journal of Control}, vol.~20, no.~1,
  pp. 24--37, 2014.

\bibitem{hoshino21}
H.~Hoshino, ``Economic benefits of solar photovoltaics and battery systems at
  customer side studied via screening curve method,'' \emph{IEEJ Transactions
  on Power and Energy}, vol. 141, no.~3, pp. 247--254, 2021, (in Japanese).

\bibitem{hoshino22:nolta}
------, ``Screening curve method for optimal sizing of photovoltaic and battery
  storage systems for a household,'' in \emph{2022 International Symposium on
  Nonlinear Theory and Its Applications (NOLTA2022)}, 2022, pp. 462--465.

\bibitem{iec61724}
{International Electrotechnical Commission}, ``Standard {IEC} 61724-1:2021:
  Photovoltaic system performance - part 1: Monitoring,''
  \url{https://webstore.iec.ch/publication/65561}.

\bibitem{pearsall2017}
N.~Pearsall, ``Introduction to photovoltaic system performance,'' in \emph{The
  Performance of Photovoltaic (PV) Systems}, N.~Pearsall, Ed.\hskip 1em plus
  0.5em minus 0.4em\relax Woodhead Publishing, 2017, pp. 1--19.

\bibitem{database}
{National Research Panel on Energy Consumption in Households}, ``Household
  energy consumption database,''
  \url{http://tkkankyo.eng.niigata-u.ac.jp/HP/HP/database/index.htm} (accessed
  6/3/2023).

\bibitem{metpv20}
{NEDO}, ``Solar radiation database,''
  \url{https://www.pvsyst.com/help/nedo-solarradiationdb.htm} (accessed
  6/3/2023).

\bibitem{telaretti16}
E.~Telaretti, M.~Ippolito, and L.~Dusonchet, ``A simple operating strategy of
  small-scale battery energy storages for energy arbitrage under dynamic
  pricing tariffs,'' \emph{Energies}, vol.~9, no.~1, pp. 12--1--20, 2016.

\bibitem{zurfi17}
A.~Zurfi, G.~Albayati, and J.~Zhang, ``Economic feasibility of residential
  behind-the-meter battery energy storage under energy time-of-use and demand
  charge rates,'' in \emph{2017 IEEE 6th International Conference on Renewable
  Energy Research and Applications (ICRERA)}, 2017, pp. 842--849.

\bibitem{maeyaert20}
L.~Maeyaert, L.~Vandevelde, and T.~D\:{o}ring, ``Battery storage for ancillary
  services in smart distribution grids,'' \emph{Journal of Energy Storage},
  vol.~30, p. 101524, 2020.

\end{thebibliography}
\bibliographystyle{IEEEtran}

\end{document}